\documentclass[
    ,final            
  ]
  {aipproc}

\usepackage{float}				
\usepackage{graphicx}			

\layoutstyle{6x9}

\begin{document}

\vspace*{-6ex}

\title{Origin of the bright prompt optical emission in the naked eye burst}


\classification{98.70.Rz}
\keywords      {$\gamma$-ray sources; $\gamma$-ray bursts}

\author{R. Hasco\"{e}t}{
  address={UPMC Univ Paris 06, UMR 7095, Institut d'Astrophysique de Paris, F-75014, Paris, France},
  altaddress={CNRS, UMR 7095, Institut d'Astrophysique de Paris, F-75014, Paris, France},
  email={hascoet@iap.fr}
}

\author{F. Daigne\thanks{Institut Universitaire de France}~}{
  address={UPMC Univ Paris 06, UMR 7095, Institut d'Astrophysique de Paris, F-75014, Paris, France},
  altaddress={CNRS, UMR 7095, Institut d'Astrophysique de Paris, F-75014, Paris, France},
  email={daigne@iap.fr}
}

\author{R. Mochkovitch}{
  address={UPMC Univ Paris 06, UMR 7095, Institut d'Astrophysique de Paris, F-75014, Paris, France},
  altaddress={CNRS, UMR 7095, Institut d'Astrophysique de Paris, F-75014, Paris, France},
  email={mochko@iap.fr}
}

\begin{abstract}
The huge optical brightness of GRB 080319B (the ''Naked Eye Burst'')
makes this event really challenging for models of the prompt GRB emission. 
In the framework of the internal shock model, we investigate a scenario where the dominant radiative process is synchrotron emission and the high optical flux is due to the dynamical properties of the relativistic outflow: if the initial Lorentz factor distribution in the jet is highly variable, many internal shocks will form within the outflow at various radii. The most violent shocks will produce the main gamma-ray component while the less violent ones will contribute at lower energy, including the optical range.
\end{abstract}

\maketitle
\vspace*{-6ex}


\section{Introduction}
\vspace*{-1ex}

Due to a fortunate chain of events, GRB 080319B was observed in the $\gamma$-ray and the optical domains with a high temporal resolution, during the whole prompt emission \citep{racusin:2008}. It peaked at a visual magnitude V=5.3, making in principle this event visible with the naked-eye, despite its cosmological distance (z=0.937).
To explain this huge optical brightness, different scenarios have already been proposed. The broadband spectrum could be produced by the Synchro-Self Compton (SSC) mechanism \citep{racusin:2008,panaitescu:2008,kumar:2009}, or  by the Synchrotron emission from two distinct electron populations in the same emitting region \citep{zou:2009}. We have investigated these two possibilities in the framework of the internal shock model, using a detailed radiative code for the emission from shock accelerated electrons \citep{hascoet:2010}. Both scenarios face severe difficulties (energy crisis, self-absorption frequency above the optical), as also suggested by other studies \citep{zou:2009}.
We present in these proceedings a new scenario where the optical brightness is produced by the synchrotron emission of a highly variable relativistic outflow, as also suggested by \citep{yu:2009}. \\
\vspace*{-4ex}

\begin{figure}[t]
\begin{tabular}{cc}
\includegraphics[scale=0.3]{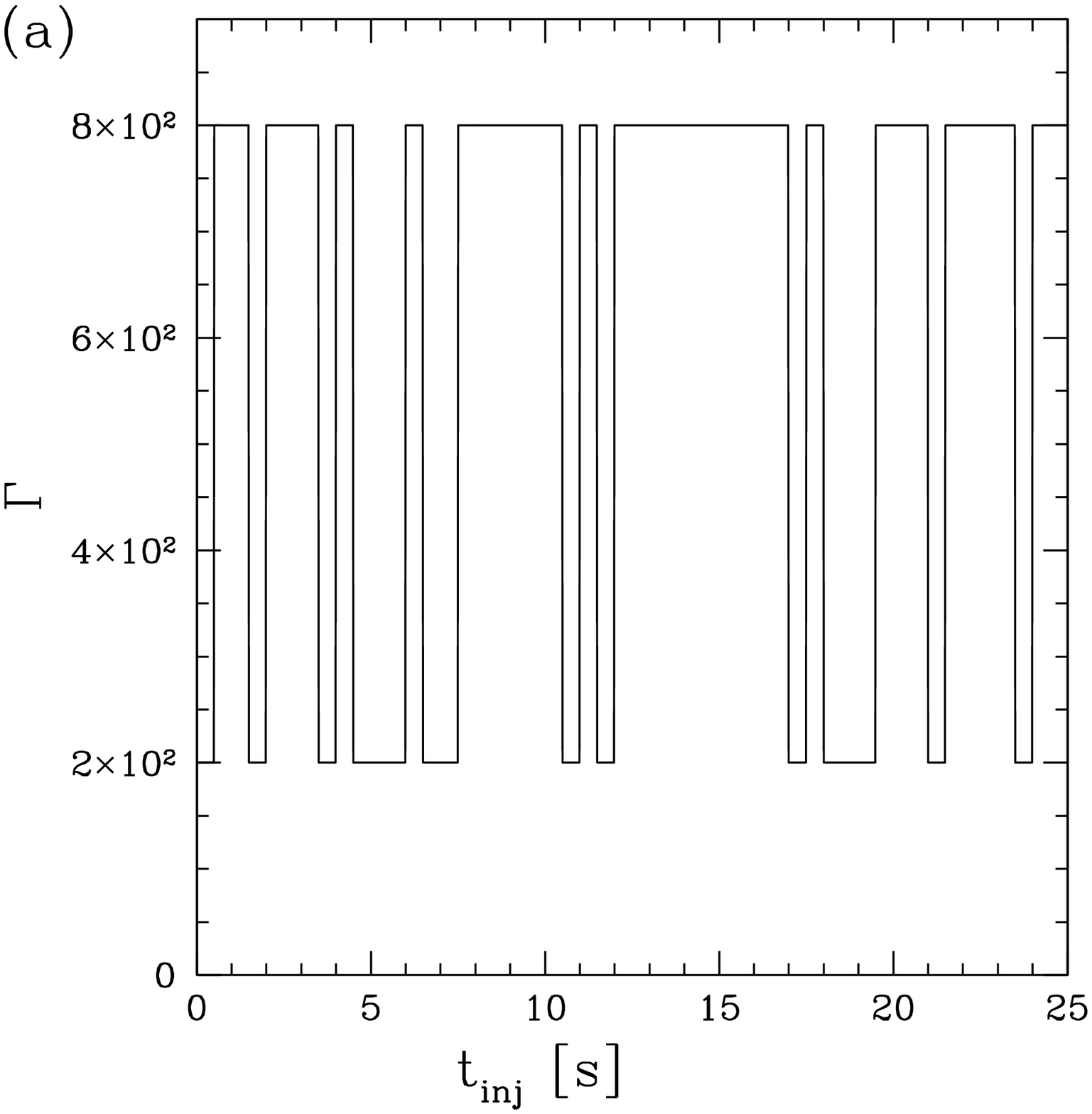} & \includegraphics[scale=0.3]{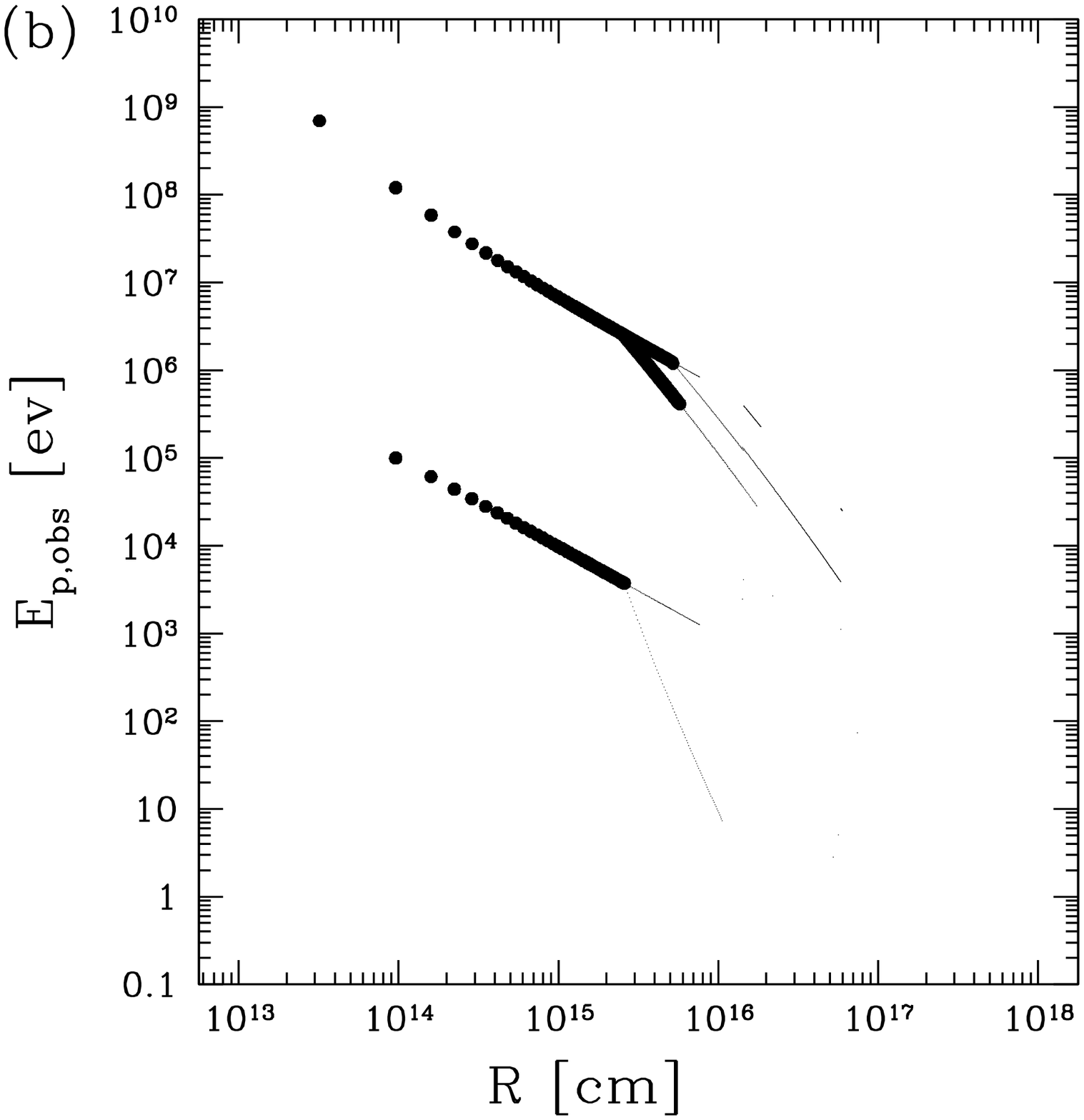}  \\ 
\includegraphics[scale=0.3]{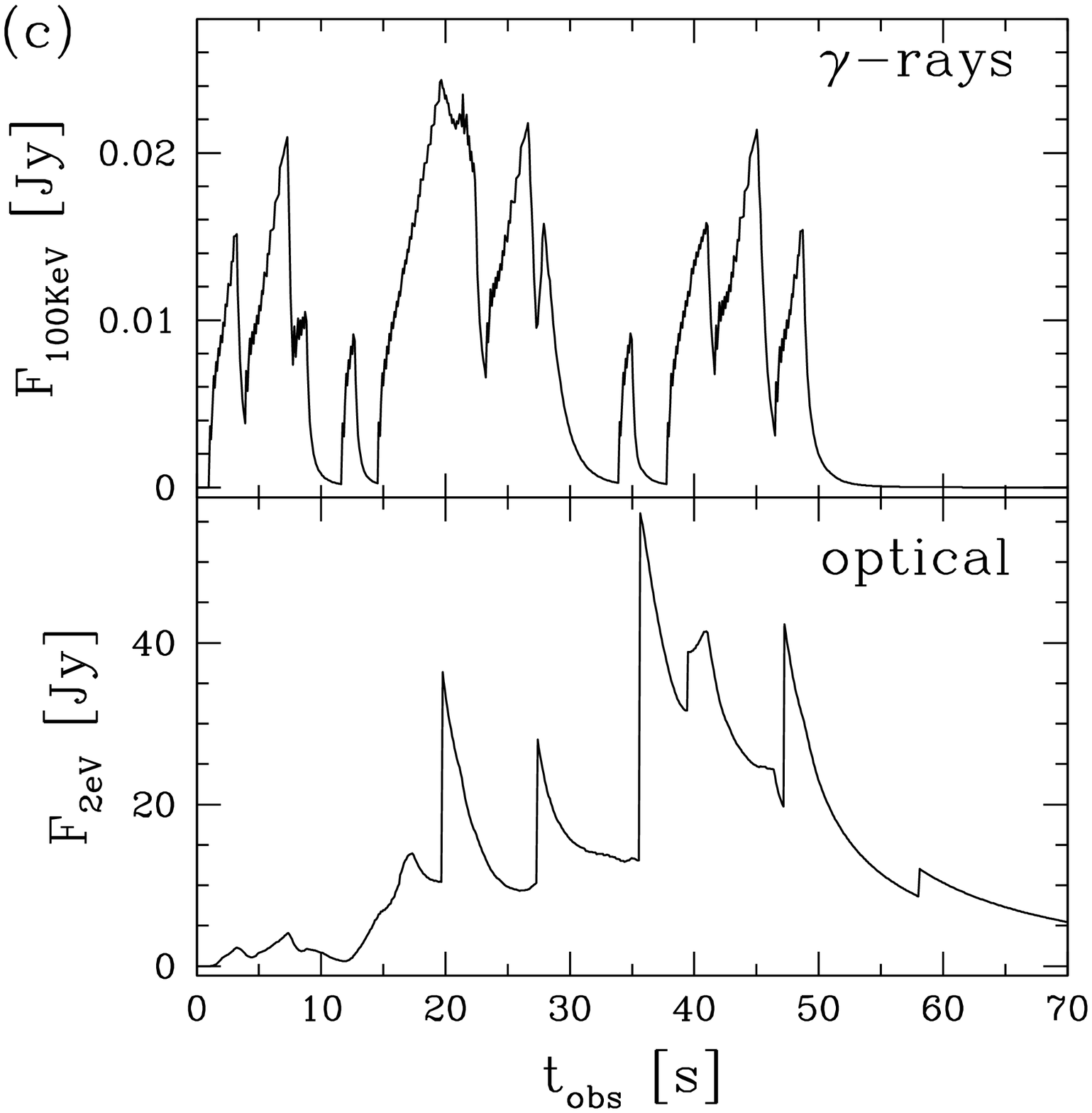} & \includegraphics[scale=0.3]{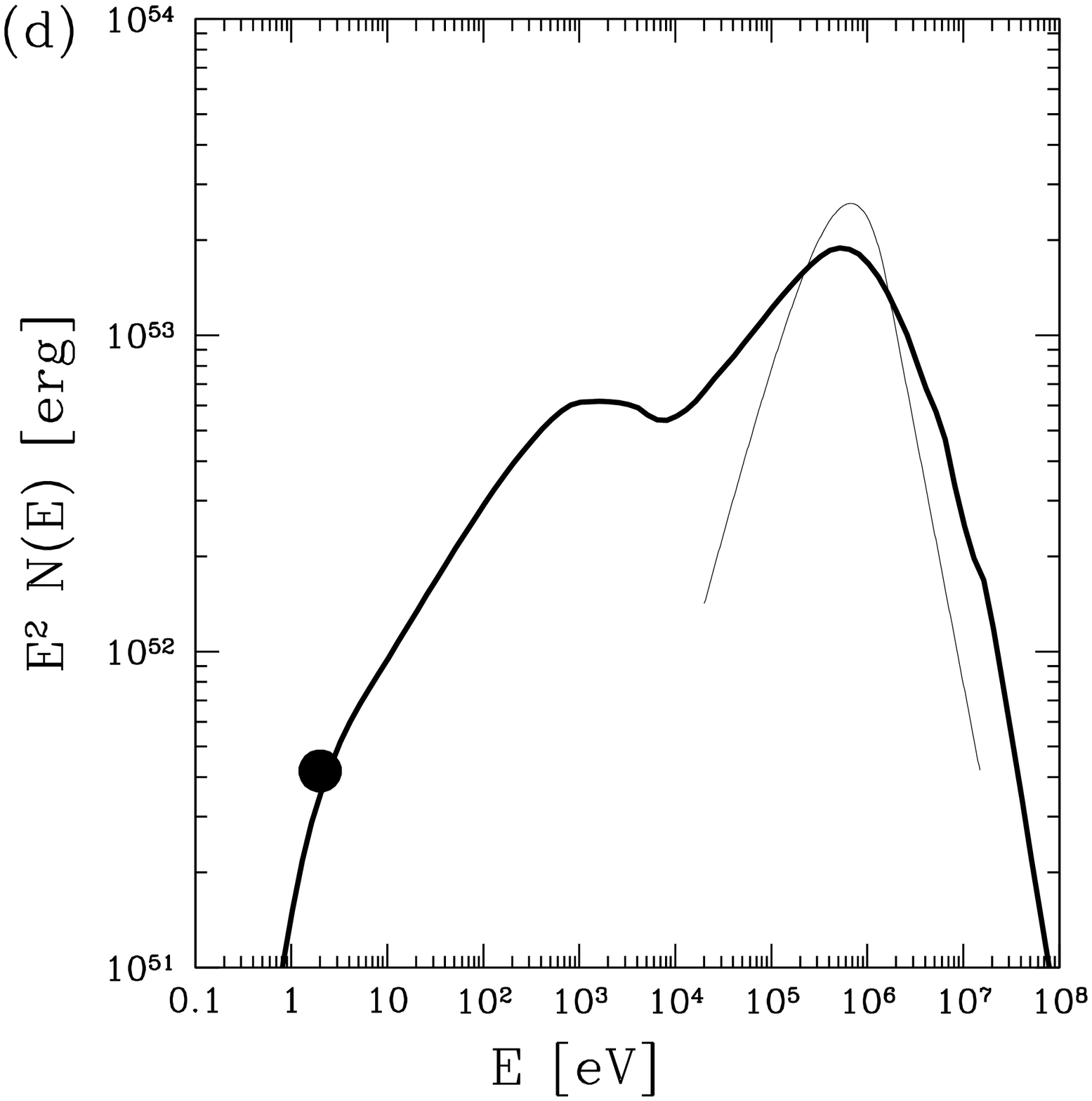}
\end{tabular}
\caption{\textbf{An example of a synthetic GRB with a bright optical flux.}
Panel (a): initial distribution of the Lorentz factor in the outflow. 
Panel (b): observed peak energy of the spectrum radiated by each collision, as a function of radius. Bigger dots indicate that the self-absorption frequency is above 2 eV.
Panel (c): light curves in $\gamma$-rays (top) and optical (bottom).
Panel (d):  simulated time-integrated spectrum (thick line), observed spectrum (thin line), observed optical level (big dot).
}
\label{good_case}

\end{figure}

\section{Bright optical emission from a highly variable relativistic outflow}
\vspace*{-1ex}

Using a simplified approach of the dynamics of internal shocks \citep{daigne:1998} and the standard prescription for the synchrotron radiation \citep{sari:1998}, we simulate synthetic GRBs produced by highly variable jets.
Fig.\ref{good_case} presents an example whose properties are very similar to those of GRB 080319B. During the relativistic ejection by the central engine, the isotropic equivalent kinetic energy injection rate is assumed to be constant ($\dot{E}_\mathrm{kin} = 2\cdot 10^{54}~\mathrm{erg\cdot s^{-1}}$), and the Lorentz factor $\Gamma$ varies on a timescale of 0.5 s. In this example, for an illustrative purpose, it is forced to be either 200 or 800 with equal probability (Fig.1a). The total ejection lasts for 25 s which leads to an observed burst duration of the order of 50 s at $z\sim 1$. Two series of internal shocks occur, which contribute either in soft $\gamma$-rays or at lower energy, depending on the violence of the collision (Fig.1b). For collisions at small radii, synchrotron self-absorption suppresses the contribution in the optical range. The predicted lightcurves show several features observed in GRB 080319B: compared to $\gamma$-rays, the optical lightcurve is less variable, and starts and ends with a delay (Fig.1c). The spectrum clearly shows the two different components associated to the two series of shocks (Fig.1d). However the low-energy slope $\alpha$ in the $\gamma$-ray range does not reproduce the observed value. This is a well known problem, which is not specific to the naked eye burst. It probably indicates that the expected prediction of the fast cooling synchrotron spectrum $\alpha=-1.5$ \citep{sari:1998} is too simple. Several effects, such as inverse Compton scatterings in Klein-Nishina regime \citep{daigne:2010}, can steepen the spectrum.

\section{Discussion and conclusion}

To test the robustness of the proposed scenario, we performed a set of simulations, where several hundreds of synthetic bursts were generated with different realizations of the initial distribution of the Lorentz factor, assuming that it varies on a timescale of 0.5 s taking random values uniformly  distributed between 200 and 800. The kinetic energy flux is adjusted in each case to  keep the radiated energy in $\gamma$-rays constant (and similar to the observed value in GRB 080319B).
This Monte Carlo analysis shows that a highly variable outflow has a fair probability of producing a bright optical emission during the prompt phase that reaches a flux above 15 Jy at a frequency of 2 eV (25 \% of the synthetic bursts). In a large fraction of this sample, the synthetic GRBs reproduce the following features: 
(i) a high optical flux. It is mainly built up by the milder internal shocks;
(ii) the optical light curve is less variable than the $\gamma$-ray one. Even if each collision radiates on a very short timescale, the corresponding contribution is detected by the observer on a timescale $\Delta t_{obs} \approx R/2\Gamma^{2}c$ due to the curvature of the emitting surface. The mild collisions dominant in the optical range have lower Lorentz factor and therefore larger $\Delta t_{obs}$. In addition, the spectral evolution during the high latitude emission is also more favorable for these mild collisions;
(iii) the optical light curve begins and ends after the $\gamma$-ray one. The delay at the beginning is due to a strong self-absorption for the first collisions with smaller radii whereas the persisting optical emission at the end is due to the high latitude emission (same reason as above) and to the fact that late collisions tend to be less violent and radiate at low energy. \\

This scenario does not predict a systematic bright optical emission for all GRBs, even in cases where the distribution of the Lorentz factor in the jet is highly variable. To better predict the frequency of events such as GRB 080319B, more details about the central engine would be needed to put more constraints on the distribution of the dynamical properties of relativistic outflows. More systematic observations of GRB prompt optical emission would be highly valuable to provide observational constraints.


\begin{theacknowledgments}
This work is partially supported by a grant from the French Space Agency (CNES). \
R.H. is funded by the research foundation from ``Capital Fund Management''. 
\end{theacknowledgments}

\bibliographystyle{aipproc}   
\bibliography{hascoet}

\end{document}